\documentstyle[12pt,epsf]{article}

\def\V{\bf V}
\def\t{\bf t}

\begin{document}
\begin{center}

\def\V{\bf V}
\def\t{\bf t}

\null\vspace{2cm}
{\large\bf KINETICS PROPERTIES OF VOLTAGE INDUCED COLICIN Ia CHANNELS
INTO A LIPID BILAYER}\\
\vspace{2cm}
{\bf R. C\'assia-Moura$^1$ and A. Popescu$^2$\\
The Abdus Salam International Centre for Theoretical Physics
Trieste, Italy\\[1cm]
1. Author to whom correspondence should be addressed at\\ permanent address:\\
Universidade de Pernambuco, Instituto de Ci\^encias Biol\'ogicas,\\
DCF-Biofisica, Caixa postal 7817,\\ 
Recife,  PE,  50732-970,  Brazil\\
E-mail address: rita@npd.ufpe.br\\ [0.5cm]
2. Permanent address: Department of Biophysics,\\ Faculty of Physics, University of Bucharest,\\
P.O. Box MG 11, Bucharest, 76 900 Romania\\
E-mail address: aure@scut.fizica.unibuc.ro}
\end{center}
\vspace{1cm}
\centerline{\bf Abstract}
\bigskip

The activation kinetics of the ion channels formed by colicin Ia incorporated into a planar bilayer 
lipid membrane (BLM) was investigated by the voltage clamp technique using different step voltage
stimuli.   The temporal behaviour of ion channels put in evidence a gain or a loss of memory, revealed by
a specific sequence of electrical pulses used for stimulation.
\vspace{1.5cm}


\newpage

\section{INTRODUCTION}

Planar bilayer lipid membrane (BLM) has been used extensively as a model of biomembranes (Taylor and
Schultz 1996, Tien and Ottova 1998).  This artificial membrane is a self-assembling system {\em in vitro},
which constitutes the fundamental spatial structure of all plasma membranes.  In fact, the lipid matrix
offers the structural frame in which all the membrane proteins are inserted (i.e., enzymes, transporters,
specific receptors, etc.).  The advantage of the BLM is that  both sides of membrane can  easily be
altered and probed by electrodes.  For long-term investigations and technological applications, planar
BLMs can also be made on either gel or metallic supports (Tien and Ottova 1998).
	
In our experiments, we incorporated colicin Ia into planar BLM.  Colicin Ia is a protein toxin secreted
{\em by Escherichia coli}, which kills the other competing bacteria (Konisky 1982).  It forms
voltage-gated ion-conducting channels both in the inner membrane of target bacteria and in planar BLM
(Kienker {\em et al} 1997).  Colicin Ia is a protein belonging to the class of bacterial toxins that share
a common strategy: they are inserted into the membrane of the target cells, punching huge holes into
them.  {\em In vivo} these holes allow the entrance of foreign particles and the escape of intracellular
components.  The dramatic exchange of charged particles have, as a result, the loss of electrochemical
membrane potential and the cellular death.  
	
Colicin Ia is a protein of 626 amino acid residues rich in charged residues (conferring to it
hydrophylicity), except for a hydrophobic segment of 40 residues near the carboxyl terminus (Wiener {\em
et al} 1997, Kienker {\em et al} 1997).  Maybe the amino acid sequence, rich in charged residues, is
necessary in the earlier stage of attachment of the colicin Ia molecule to the membrane, whose external
surface is also hydrophylic.  Then colicin Ia binds with the hydrophobic segment parallel  to the
membrane and this portion is inserted in a transmembrane orientation (Kienker {\em et al} 1997).  Colicin
Ia translocates a large hydrophilic part of itself completely across a lipid bilayer in conjunction with
the formation of a ion-conduction channel (Jakes {\em et al} 1998).  At least 68 residues flip back and
forth across the membrane in association with channel opening and closing (Qiu {\em et al} 1996). 
Channel-forming colicin contacts the inner and outer membranes simultaneously during function (Qiu {\em et
al} 1996).
	
The probability of the opening of voltage-gated ion channels is determined by the transmembrane potential
(Tzounopoulos {\em et al} 1998).  This channel activation results from a series of conformational
transitions in the channel protein, particularly the movement of charged residues within the membrane
electric field (Armstrong and Bezanilla 1973).  In this work an electrophysiological method is used in
order to demonstrate that it is possible to control the activation kinetics of the colicin Ia
incorporated into planar BLM.  In this way it was found experimentally that this artificial membrane (BLM
+ colicin Ia) can gain or can lose its memory.


\section{MATERIAL AND METHODS}

The experimental procedure was according to the method previously described (C\'assia-Moura 1993).  
The lipid bilayer was formed by opposing two lipid monolayers across a small hole of 180-200 $\mu m$
diameter, situated in a Teflon wall separating two aqueous solutions.  Lipid monolayers were formed using
azolectin (L-$\alpha$-phosphatidylcholine type II - Sigma Chemical Co) dissolved in hexane, so that
solutions of 1\% concentration were realised.  During bilayer formation the peak current as response to
constant voltage stimulation (amplitude: $\pm 10$ mV; duration: 2-4 ms; frequency: 500 Hz) was
continuously monitored.
	
The aqueous solution consisted of 500 mM KCl+5 mM CaCl$_2$+5mM HEPES+1mM EDTA (final pH 7.00).  Deionized
water, double distilled in glass, was used in the preparation of all solutions.  All chemicals were of
analytical grade.
	
Colicin Ia was added directly to the aqueous solution in one half of the device realising a final
concentration of 1-5 g/ml.  For the sake of clarity, we shall refer to the side of the BLM containing
colicin Ia as to a {\em cis} half, while the colicin-free side will be called {\em trans} half.  Two
Ag/AgCl electrodes were used to connect the electronics to the solutions (one electrode in each half) via
salt bridges (2.5\% agar in the chamber medium, electrodes immersed in 3 M KCl).  A pulse generator with a
d.c. voltage of $\pm  200$ mV was connected to the {\em cis} half.  The ionic currents flowing through the
artificial membrane were measured under conventional voltage-clamp conditions using an operational
amplifier (Burr-Brown model OPA111) in the current-to-voltage converter configuration.  The converter
input was connected to {\em trans} half, and the output was connected to a physiograph (Gilson) either
directly or via a digital oscilloscope (Nicolet model 201).  Experiments were performed at room
temperature, that is at $25\pm 2$ $^\circ$C.


\section{RESULTS AND DISCUSSION}

More than 2500 recordings of ionic current versus time were obtained from 40 artificial membranes and
analysed in this work.  The current was produced by pulses of at least 20 s duration of the step voltage
stimulus.  We do not intend to present all these crude experimental data, but merely their
common aspects.  The reason is that we noticed a variability of the artificial membrane behaviour, in
spite of the fact that the experimental procedure was strictly respected in all the cases.
	
Although it was observed a slight different quantitative current response of the artificial membranes to
the applied voltage pulses, however one can remark, in all cases, the following common features:
\begin{itemize}
\item 
a lack of response to negative pulses (Figure 1a and 1b);
\item 
an inconsistent response to positive pulses less than 50 mV;
\item  
a brief response and an irreversible breakdown of membrane structure to positive pulses greater
than 90 mV;
\item  
a ``window''  of positive pulses  $(\V)$  between  50 mV and 90 mV, in which all records
obtained with the same positive pulse presented a sudden exponential rise of the ionic current and a
consecutive longer linear increase phase (Figure 1: $R_1 - R_6$).  On the other hand, some of them were
different both in the amplitude and time course.  At this  ``window'', a different pattern of responses
can be observed for any single voltage value applied to the artificial membrane, but a single stimulus
always produce a unique response.  The variability in this case is the result of successive repetitive
trials on the same preparation, of different measurements on several preparations or of the large time
intervals between the measurements.  The time required to reach the end of the exponential phase varied
from 7 to 50 s.
\end{itemize}
	
As concerning  the repetitive stimulations, we have put in evidence a critical interval, $\Delta\t_c$,
between two identical successive pulses $\V$ interposed by a resting period $(\Delta t)$, with the
following characteristics:
\begin{itemize}
\item
it is less than 120 s  and is specific to each particular membrane;
\item  
the second pulse applied within $\Delta\t_c$ always performs the same current response if the resting
period is maintained.  Moreover, the response manifests a greater amplitude if the resting period between
the pulses is smaller (Figure 1a: $R_1 - R_3$);
\item 
the second pulse applied after $\Delta\t_c$ always conducts to a non controlled current response, both in
the exponential phase as in the linear one (Figure 1b: $R_4 - R_6$).
\end{itemize}
	
The results of this study were surprising in at least one aspect:  namely, from Figure 1 one can
observe that if  $\Delta t_1  < \Delta t_2 < \Delta t_3 < \Delta\t_c$, then the responses are decreasing
in the order  $R_1, R_2, R_3$, while if  $\Delta t_4 > \Delta\t_c$  the response is no more unique.  
	
On the other hand, from  Figure 2, it results that when we apply a sequence of two successive pulses
($P_1$ and $P_2$) separated by a resting period within $\Delta\t_c$ $(\Delta t_{1-2} < \Delta\t_c)$,
followed by a third one ($P_3$) separated by a resting period after $\Delta\t_c$ $(\Delta t_{2-3} > \Delta
\t_c)$, the system will respond stochastically ($R_3$).  If  by chance the response $R_3$ is identical to
$R_1$ (see bold $R_3$), then the application of a fourth pulse ($P_4$) identical to the second one ($P_2$)
and after the same resting period before $P_2 (\Delta t_{3-4} = \Delta t_{1-2})$, the last one will induce
a response identical to that elicited by the $P_2$ pulse.  Therefore, the experimental model is
``remembering'' the previous state (installed after the $P_1$ action).  Moreover, this state can be
induced by a different pulse but pertaining to the mentioned  ``window''.  This is a proof of the memory
of this system.  The memory here must be interpreted in the general sense: if the actual state of a system
depends on its previous states, then this system is endowed with memory.
	
In the case of the ``window'' pulses, applied into an interval smaller than $\Delta\t_c$, the enhancement
(see Figures 1 and 2) of response can be attributed to the ``memory'' of the artificial membrane.  

	
The interpretation of this artificial membrane behaviour to voltage pulses is a very difficult task. 
However, one could advance some hypotheses concerning this complex behaviour.
\begin{itemize}
\item   
After the stimulation of the artificial membrane, some colicin Ia molecules, which are forming
membrane channels do not have  enough time to relax themselves (i.e., to perform a transition from the
open state to a complete closed one).  But, because the order of magnitude of $\Delta\t_c$ is too high as
compared with the time of allosteric transition of protein, one could advance the additional hypothesis
that the lipid frame of the artificial membrane will prevent a fast transition of these molecules between
the two states (open and closed).  As we already presented, at least 68 amino acid residues flip back and
forth across the membrane.
\item   
Another possibility could be of the increase the density of colicin Ia incorporated into the
artificial membrane under the influence of the electric pulses themselves.
\item   
It could also be possible that the voltage pulses induce some colicin Ia molecule aggregation,
resulting thus some clusters formation with a higher ionic conductivity than that of the individual
channels.
\end{itemize}
	
As concerning the stochastic behaviour of the artificial membrane in response to  ``window'' pulses, but
applied at intervals longer than $\Delta\t_c$, up to now we do not have a plausible explanation.  Perhaps
one could interpret this strange behaviour also in the term of ionic channels' memory of a long term type.
	
All these hypotheses must be confronted in the future with new experimental data.  We intend to extend
this study to BLM with different chemical compositions and also to biological membranes, in order to have
a better understanding of the complex bioelectrogenesis of the specialised tissues (e.g. heart, brain,
etc.) and to furnish data for an improved electrodiagnosis and electrotherapy.



\section{CONCLUSIONS}
	
We used a quite simple model of biological membranes namely BLM with colicin Ia incorporated, in order to
test its kinetics properties.  By an electrophysiological method, we demonstrated that it is possible to
control the activation kinetics of the experimental model.  In this model the ion channel memory is
experimentally induced and can be controlled by a specific sequence of pulses used for electrical
stimulation.
\vspace{1cm}

\noindent{\bf Acknowledgements}

This work was concluded at and partially supported by the Abdus Salam International Centre for
Theoretical Physics, during the authors' visit under Associate Scheme. 
	
C\'assia-Moura wishes to thank to Prof. Emanuel Dias for his constructive comments; and also  C\'esar A
S Andrade, Jos\'e Ricardo S A Lima, Katarine S A Lima and Luciana S Ventura for their stimulating
discussions and valuable technical assistance.
	
The experiments were partially performed at Departamento de Biofisica/UFPE.  A part of this work was
supported by grants from CNPq and CAPES.

\newpage

\noindent{\bf REFERENCES}
\begin{description}
\item{}
Armstrong C M and Bezanilla F 1973 Movement of sodium ions associated with the nerve 	impulse
{\em Nature} {\bf 242} 457-461 
\item{}
C\'assia-Moura R 1993 Activation kinetics of the incorporation of colicin
Ia into an artificial  membrane: a Markov or a fractal model? {\em Bioelectrochem. Bioenerg.} {\bf 32}
175-180 
\item{}
Jakes K S, Kienker P K, Slatin S L and Finkelstein A 1998 Translocation of inserted foreign 
epitopes by a channel-forming protein {\em Proc. Natl. Acad. Sci. USA} {\bf 95} 4321-4326 
\item{}
Kienker P K, Qiu X, Slatin S L, Finkelstein and Jakes K S 1997 Transmembrane insertion of 	the colicin Ia
hydrophobic hairpin {\em J. Memb. Biol.} {\bf 157} 27-37 
\item{}
Konisky J 1982 {\em Annu. Rev. Microbiol.} {\bf 36} 125 
\item{}
Qiu X Q, Jakes K S, Kienker P K, Finkelstein A and Slatin S L 1996 Major transmembrane 	movement
associated with colicin Ia channel gating {\em J. Gen. Physiol.} {\bf 107} 313-328  
\item{}
Taylor R F and Schultz J S (eds) 1996 Bilayer lipid membranes and other lipid based methods,	in Handbook
of chemical and biological sensors (Philadelphia: Institute of Physics  Publishing) 
\item{}
Tien H T  and Ottova A L  1998 Membrane biophysics: as viewed from experimental bilayer lipid membranes
\item{}
Tien H T and Ottova A L 1998 From self-assembled bilayer lipid membranes (BLMs) to supported BLMs on
metal and gel substrates to practical applications (Colloids and  surfaces) 
\item{}
Tzounopoulos T, Maylie J and Adelman J P 1998 Gating of $I_{sK}$ channels expressed in {\em	Xenopus}
oocytes {\em Bhiophys. J.} {\bf 74} 2299-2305 
\item{}
Wiener M, Freymann D, Ghosh P and Stroud R M 1997 Crystal structure of colicin Ia {\em Nature}
{\bf  385} 461-464
\end{description}

\newpage

\centerline{\bf Figure Captions}
\bigskip

\noindent FIGURE   1   -  The time course response of our experimental model as a function of the pulse
characteristics (voltage applied and resting period).  Upper curve:  the time course of step pulses
applied.\\
a)  The responses to pulses applied within $\Delta\t_c$.  In this case the responses are deterministic.\\
b)  The responses to pulses applied after a resting period greater than $\Delta\t_c$.  In this case the
responses are not predictable ($R_4, R_5, R_6$ or any other responses).\\ \\
FIGURE  2  -  The ionic current time response $(R_1 - R_4)$ of the experimental model as a consequence of
a specific temporal sequence of voltage pulses $(P_1 - P_4)$. The applied voltage $P_3$ are in the range
of $P_{3{\rm min}} = +50$ mV and $P_{3{\rm max}} = +90$ mV (that is, within the  ``windows'').  $P_1, P_2$
and $P_4$ are also in this range as $P_3$ but they are fixed, while $P_3$ has any value in this range. 
$\Delta\t_c$ is specific to each artificial membrane (for details, see the text).  Note that  $\Delta
t_{2-3} > \Delta\t_c$  and in this case, there are many uncontrolled possibilities of $R_3$ responses (but
at a given pulse a single response is obtained).  The response $R_3$ (bolded in the figure) is identical
with $R_1$ (although it is possible that $P_3$ that generates it, is not identical to $P_1$).  In this
case, the stimulation of the experimental model with a $P_4$ pulse after  $\Delta t_{3-4} < \Delta\t_c$ 
will produce the response $R_4$ identical to the response $R_2$, if the resting period $\Delta t_{1-2}$ is
equal to $\Delta t_{3-4}$, and if the applied voltage $P_4$ is equal to $P_2$.

\newpage

\begin{figure}
\begin{center}
{\parbox[t]{16cm}{\epsfxsize 16cm
\epsffile{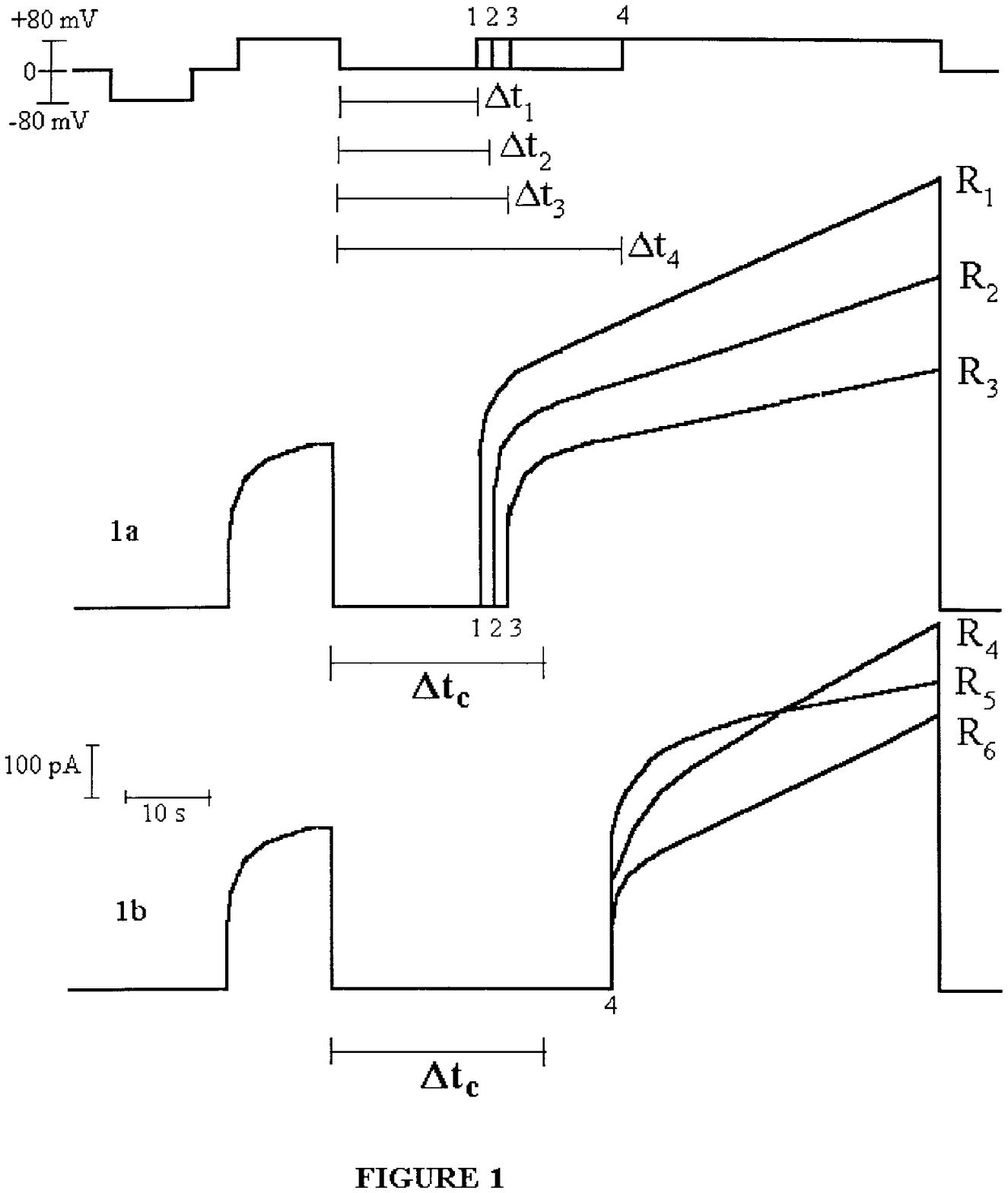}}
}
\end{center}
\end{figure}

\newpage

\begin{figure}
\begin{center}
{\parbox[t]{12cm}{\epsfxsize 12cm
\epsffile{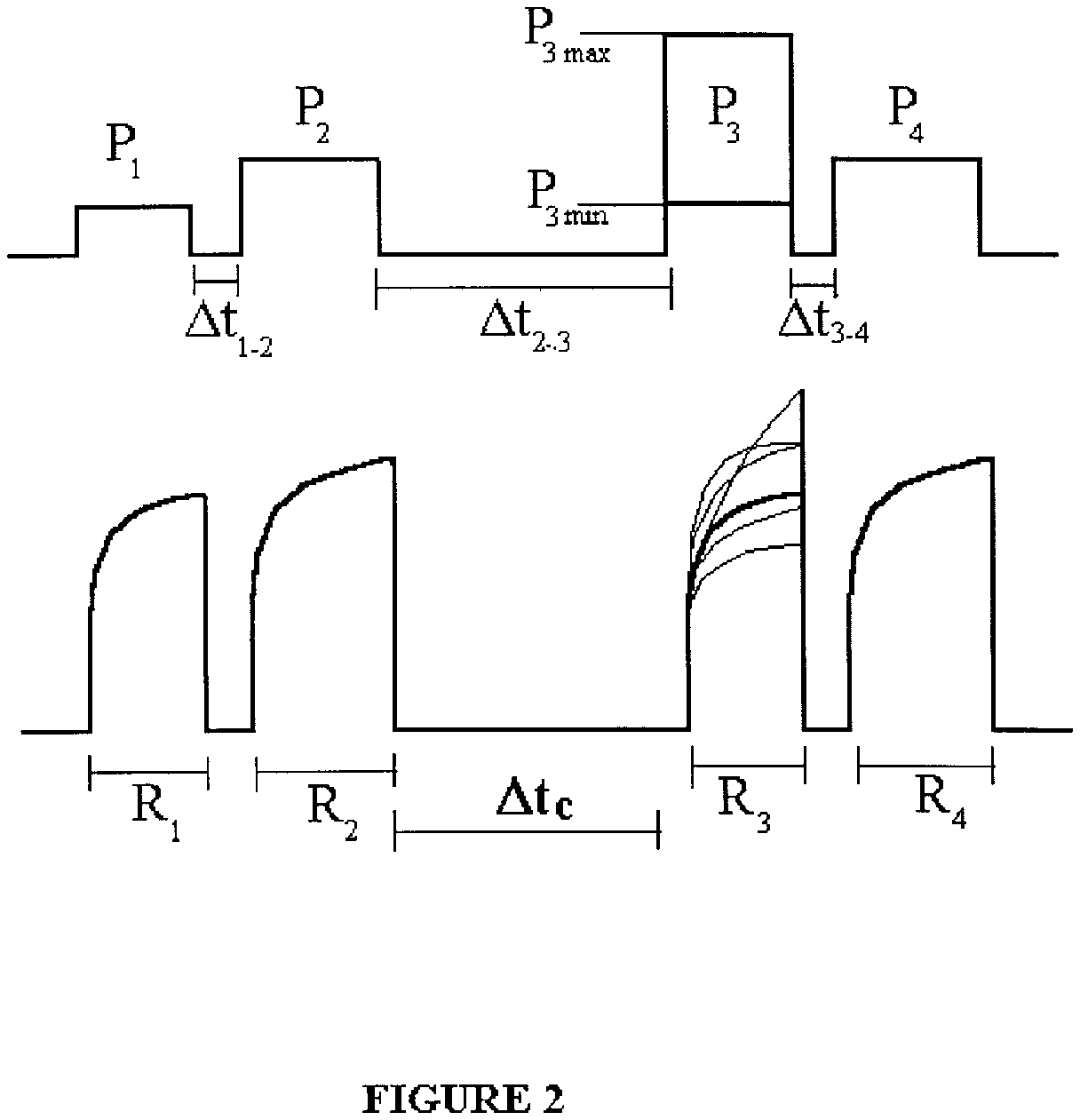}}
}
\end{center}
\end{figure}

\end{document}